\documentclass[aps,prl,reprint,groupedaddress,showpacks,showkeys]{revtex4-1}

\usepackage{amsmath,amssymb}
\usepackage{graphicx}
\usepackage{xcolor}
\usepackage{ulem}
\usepackage[export]{adjustbox}

\usepackage[version=3]{mhchem} 
\usepackage[T1]{fontenc}       

\begin{document}

\title{Coupling of Coexisting Non-Collinear Spin States in the Fe Monolayer on Re(0001)}

\author{Alexandra Palacio-Morales}
\affiliation{Department of Physics, University of Hamburg, D-20355 Hamburg, Germany}

\author{Andr\'e Kubetzka$^{*}$}
\affiliation{Department of Physics, University of Hamburg, D-20355 Hamburg, Germany}

\author{Kirsten von Bergmann$^{\dagger}$}
\affiliation{Department of Physics, University of Hamburg, D-20355 Hamburg, Germany}

\author{Roland Wiesendanger}
\affiliation{Department of Physics, University of Hamburg, D-20355 Hamburg, Germany}


\begin{abstract}
 \begin{minipage}{0.4\linewidth}   
Spin-polarized scanning tunneling microscopy is used to investigate the magnetic state of the Fe monolayer on Re(0001). Two coexisting atomic-scale non-collinear spin textures are observed with a sharp transition between them on the order of one atomic lattice spacing. A strict position correlation between the two spin states is observed both in experiments and in Monte Carlo simulations, demonstrating their strong coupling behavior.
  \end{minipage}
  \begin{minipage}{0.4\linewidth}
\includegraphics[height=3.5cm, left]{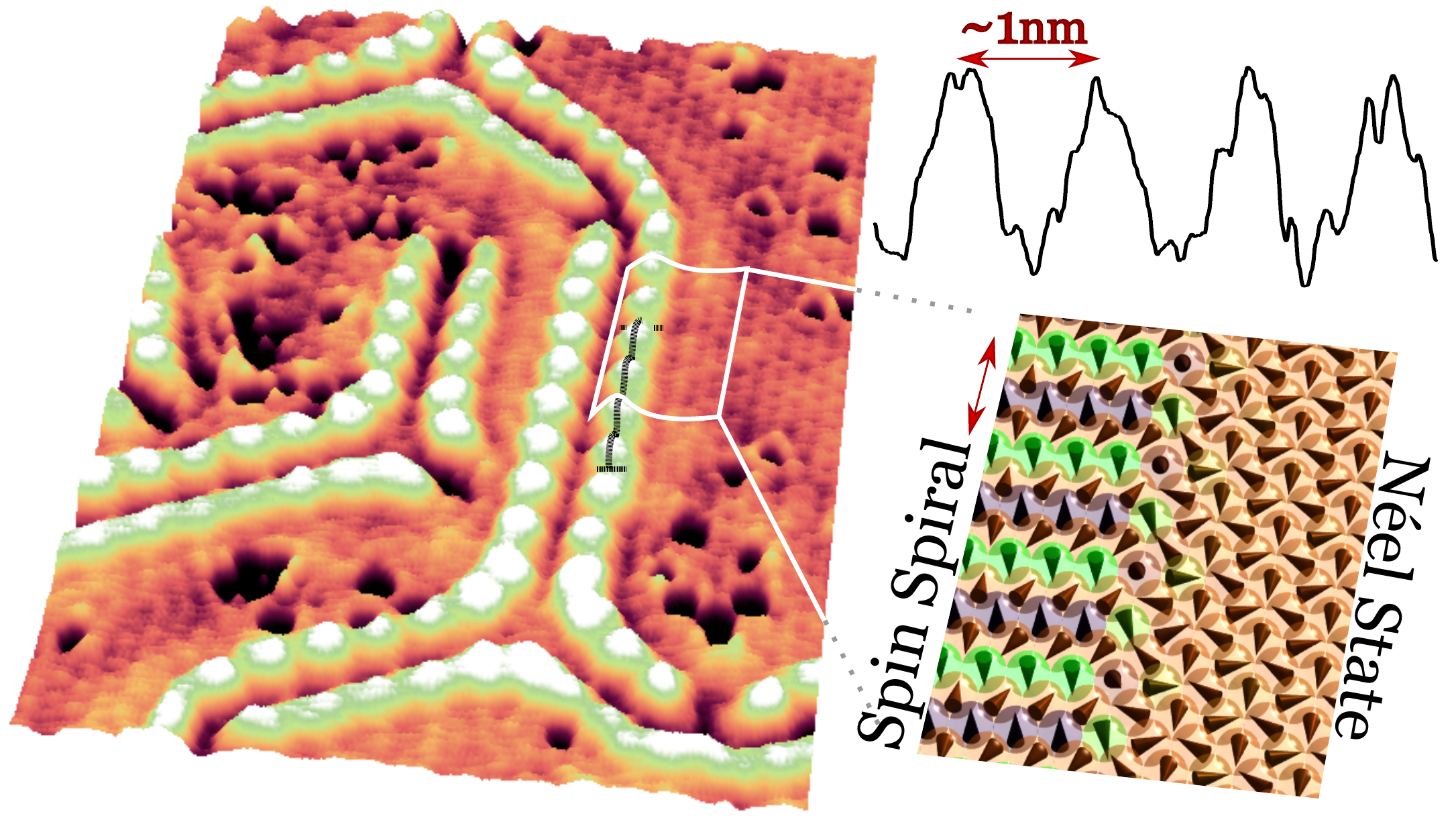}
  \end{minipage}
\end{abstract}


\maketitle

Recently, there has been increasing interest in non-collinear magnetic states in ultrathin films and several complex spin textures have been reported~\cite{vonBergmannJPCM2014}. Typically such non-collinear spin configurations arise due to competing interactions, for instance nearest neighbor and next-nearest neighbor Heisenberg exchange, which can lead to spin spirals with a period given by the ratio of the interaction strengths. When the Dzyaloshinskii-Moriya interaction (DMI) is involved, such spin spirals exhibit unique rotational sense~\cite{BodeNature2007,Ferriani2008} and may transform into magnetic skyrmions in external magnetic fields~\cite{Romming2013}. However, the prototype non-collinear state in two dimensions is the N\'{e}el state, which is the ground state for spins on a hexagonal lattice with antiferromagnetic nearest neighbor Heisenberg exchange interaction, and has been observed for several pseudomorphic monolayer 3$d$ transition metal films on 5$d$ substrates~\cite{Gao2008,Wasniowska2010,Ouazi2014_PRL}. The magnetic interactions of a thin-film system are governed by the distance and the number of nearest neighbors, by the adsorption geometry, and the hybridization with the non-magnetic substrate. For Fe thin films it has been demonstrated that not only different substrates~\cite{Hardrat2009,Palotas2014} but also a different stacking, uniaxial strain relief, or different symmetry of a surface can lead to a variety of magnetic ground states~\cite{Kubetzka2005,vonBergmannJPCM2014,HsuPRL2016,vonBergmannNanoLett2015b}.  This suggests that different non-collinear magnetic ground states may coexist, giving rise to the question about mutual interactions between them.

Here, we report on spin-polarized scanning tunneling microscopy (SP-STM) measurements~\cite{Wiesendanger2009} of an Fe monolayer on Re$(0001)$. In addition to the previously observed N\'{e}el state in the pseudomorphic hcp-stacked Fe monolayer~\cite{Ouazi2014_PRL} , we find a spin spiral state in dislocation lines that are incorporated to release lateral strain. A close analysis reveals that within a dislocation line the spins are canted with respect to the dislocation line propagation. At the interface between these two coexisting non-collinear magnetic states we find a sharp transition on the order of a few atoms only. In addition, a spatial correlation between the spin spiral and the N\'{e}el state suggests a coupling between the two states, which is confirmed by Monte-Carlo simulations.

\begin{figure}[]
	\includegraphics[width=8.5cm]{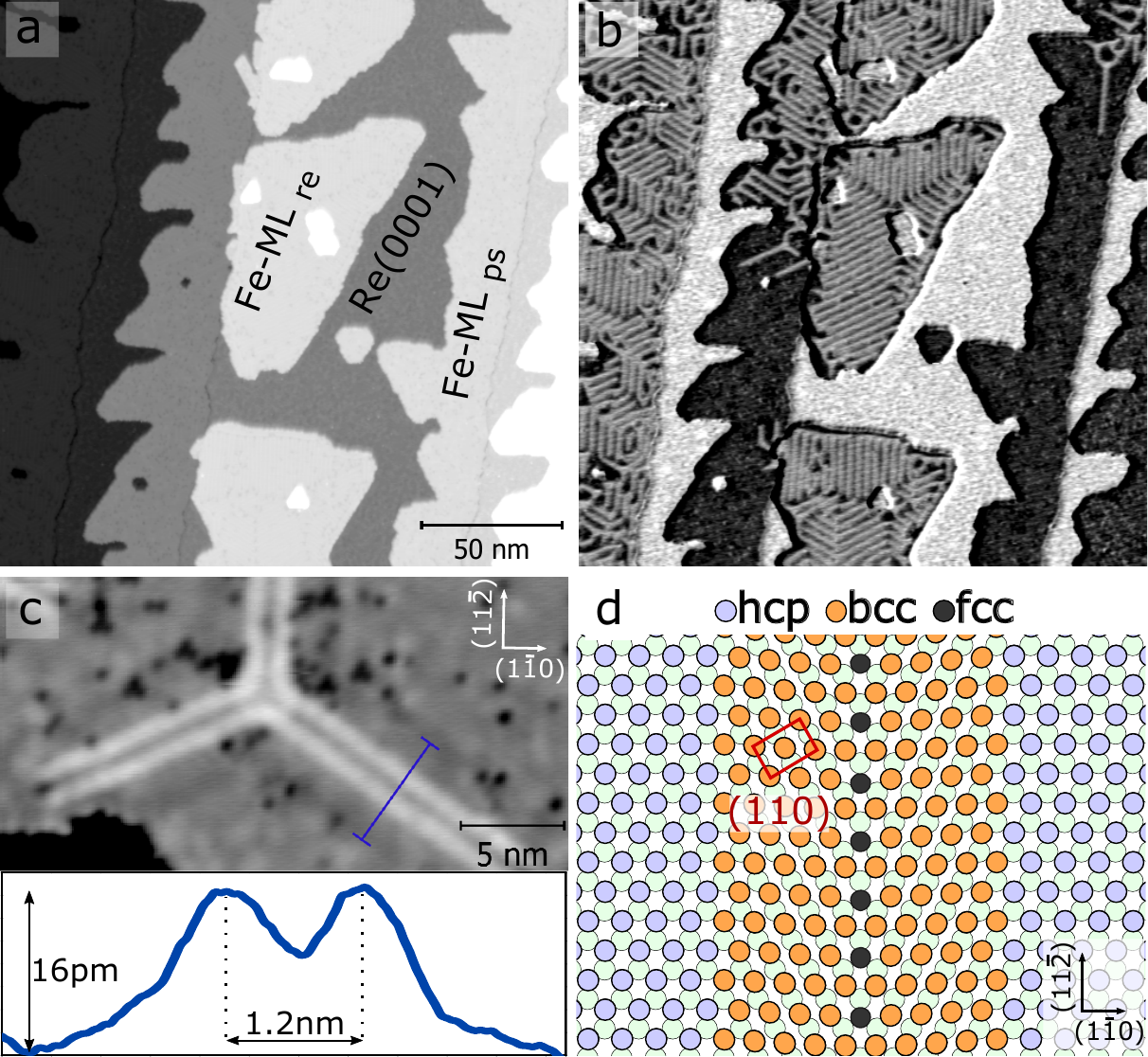}
	\caption{\label{Fig1} (Color online) \textbf{(a)}~Constant-current STM image of about $0.6$\,ML of Fe on Re$(0001)$ ($U=+100$\,mV, $I=1$\,nA, $T=4.6$\,K). The Fe-ML is either pseudomorphic (Fe-ML$_{\rm ps}$) or reconstructed (Fe-ML$_{\rm re}$). \textbf{(b)}~Simultaneously measured d$I$/d$U$~map ($U_{\rm mod}=10$\,mV). \textbf{(c)}~Constant-current STM image of single dislocation lines along $<11\overline{2}>$ and height profile along the line indicated. (d)~Atomic structure model of a dislocation line. The rectangle indicates a bcc$(110)$-like unit cell.}
\end{figure}

Figure~\ref{Fig1}(a) shows an overview STM image of submonolayer Fe growth on Re$(0001)$. The temperature during Fe deposition was around $300$\,K, which results in both step flow growth and island nucleation. As evident from the simultaneously measured map of differential tunneling conductance d$I$/d$U$ in (b), the Fe monolayer~(ML) has pseudomorphic (dark, Fe-ML$_{\rm ps}$) and reconstructed (grey lines, Fe-ML$_{\rm re}$) areas, similar to other Fe systems on hexagonal substrates~\cite{HsuPRL2016} . As reported in Refs.~\cite{Ouazi2014_PRL,Hardrat2009} 

\newpage

\onecolumngrid

\begin{figure}
	\includegraphics[width=17.7cm, left]{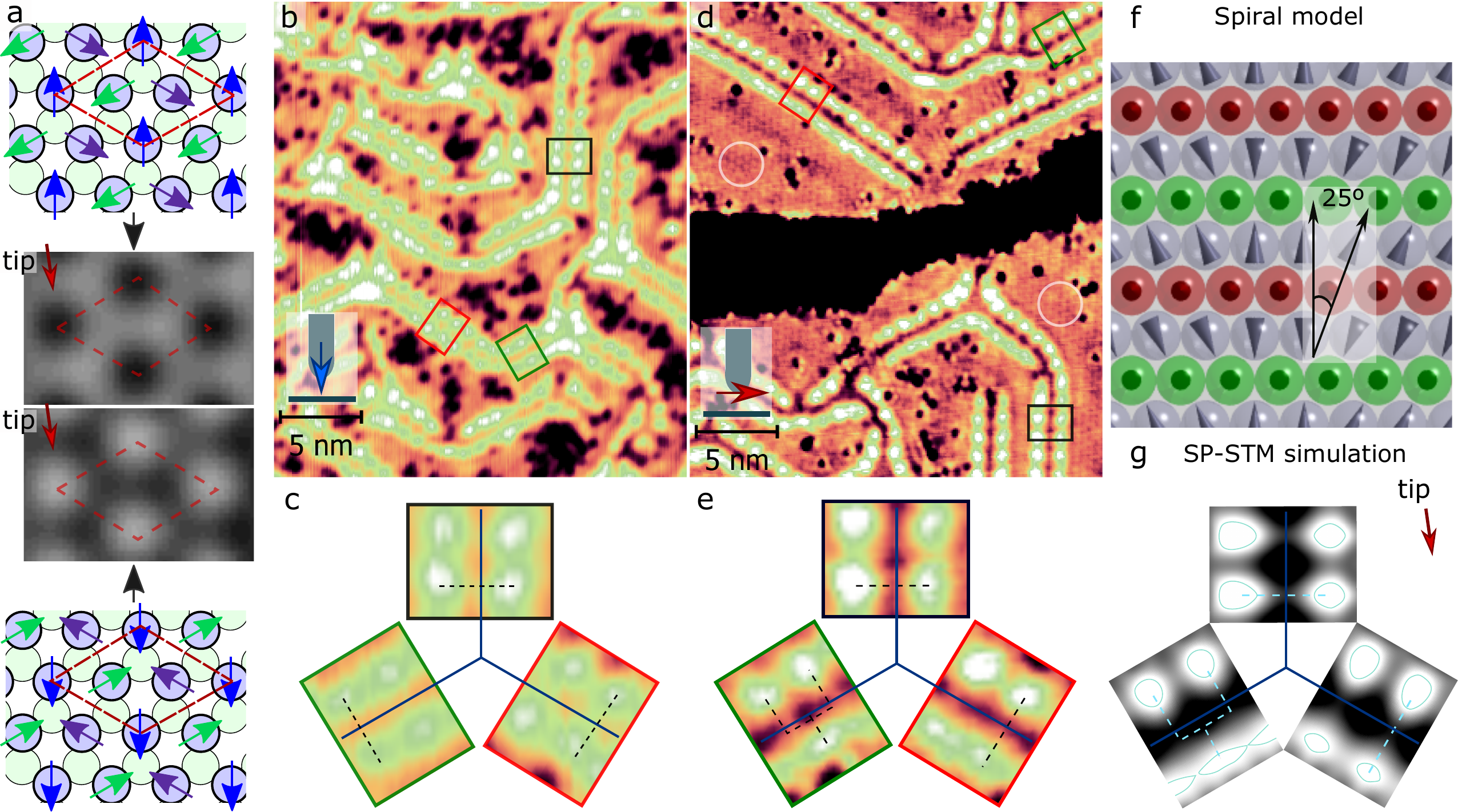}
	\caption{\label{Fig2} (Color online) \textbf{(a)}~Two magnetic N\'eel states with the same rotational sense and the corresponding SP-STM simulations for the given in-plane magnetized tip. \textbf{(b)}~Overview SP-STM constant-current image with an out-of-plane sensitive Cr tip. \textbf{(c)}~Magnified views of the boxes indicated in (b). The black dashed lines indicate the maximum contrast relation between the two strands of the dislocation lines. The solid lines show the spiral directions. \textbf{(d)}~Overview SP-STM constant-current image with an in-plane sensitive Cr tip. Two white circles indicate the two different domains of the N\'eel state coexisting on the sample. \textbf{(e)}~Magnified views of the boxes indicated in (d). The black dashed lines indicate the maximum contrast relation between the two strands of the dislocation lines and the solid line the direction of the spirals. \textbf{(f)}~Model of the spin spiral within the dislocation lines as deduced from the experimental data with cycloidal nature and spins canted $\pm 25^{\circ}$ away from the vertical. \textbf{(g)}~SP-STM simulations~\cite{Heinze2006} of the spin spiral model (f) for the three rotational symmetric dislocation lines, displayed by solid lines, for a tip magnetization as indicated (the height of the atoms was adjusted to fit the experimental height variation due to the differently stacked atoms, cf. fit to the experimental line profile in Fig.\,\ref{Fig3}(b), polarization of the tunnel current was $30\%$ and the tip-sample distance was $0.6$\,nm). (Measurement parameters for (b)-(e): $U= +10$\,mV, $I= 1$\,nA, $T=4.6$\,K).}
\end{figure}

\twocolumngrid

\noindent the Fe atoms in the pseudomorphic area reside in the hcp hollow sites with a nearest neighbor distance of 274\,pm, given by the Re substrate. The dislocation lines appear in the three equivalent $<11\overline{2}>$ crystallographic directions of the crystal and tend to accumulate to form uniaxially reconstructed areas with a minimum periodicity of about 3.2\,nm. For larger Fe coverage the amount of reconstructed areas increases~\cite{Ouazi2014_PRL}. One can also find single dislocation lines embedded in an extended pseudomorphic Fe layer, see Fig.\,\ref{Fig1}(c). They are imaged as two parallel bright strands with a distance of about 1.2\,nm. A structure model is shown in Fig.\,\ref{Fig1}(d), where 9 Fe atoms are spread over a distance of 8 Re atoms. This leads to gradual change of the Fe adsorption site from hcp over bridge sites to fcc in the center of the dislocation line. When the Fe atoms reside in bridge sites the structure resembles bcc(110) as indicated by the red rectangle. The height profile displayed at the bottom of Fig.\,\ref{Fig1}(c) is taken along the line indicated in the image and shows that the apparent height is lower in the center of the dislocation line and maximum at the bridge site area, leading to the appearance of two strands for each dislocation line.

Previous SP-STM measurements have demonstrated that the magnetic ground state of the Fe monolayer on Re(0001) is the N\'{e}el state~\cite{Ouazi2014_PRL} , in agreement with density functional theory calculations~\cite{Hardrat2009,Palotas2014} . This state emerges due to geometrical frustration of antiferromagnetic nearest neighbor Heisenberg exchange. Recent relativistic calculations including anisotropy and the DMI predict that the spins are pointing along $<11\overline{2}>$ directions and that the magnetic state occurs with unique rotational sense~\cite{Palotas2014} . The latter is due to interface-induced DM interaction, which is usually approximated to be in the plane of the interface. However, for a fully in-plane magnetized N\'{e}el state an in-plane DM vector does not select a particular rotational sense and the more refined model of a three-site hopping mechanism between adjacent Fe atoms and a connecting Re atom has to be taken into account: the DM vector is assumed to be perpendicular to the plane spanned by these three atoms, i.e.\ perpendicular to the Fe-Fe bond and alternating between $\pm 22^{\circ}$ with respect to the surface plane~\cite{vonBergmannNanoLett2015b, Fert.MaterSciForum.1990} . The two magnetic domains of an in-plane N\'{e}el state with the rotational sense as predicted~\cite{Palotas2014} are sketched in Fig.\,\ref{Fig2}(a), together with SP-STM simulations~\cite{Heinze2006} with a tip magnetization direction as indicated by the arrow.

Figure~\ref{Fig2}(b) shows an SP-STM image of the Fe monolayer on Re with several dislocation lines. No magnetic signal is observed for the pseudomorphic areas, but the two bright strands of each dislocation line show a periodic modulation originating from a spin spiral state. The magnetic period is about $1.0$\,nm, i.e.\ about 4 atomic rows (cf.\ Fig.\,\ref{Fig1}(d)). For the three possible directions of dislocation lines the magnetic contrast is the same and on the adjacent strands the maxima of the spiral are in phase, see magnified views of the boxes in Fig.\,\ref{Fig2}(c). From the absence of magnetic contrast in areas exhibiting the N\'{e}el state and the identical magnetic contrast on rotated symmetry-equivalent dislocation lines we conclude that the tip is sensitive to the out-of-plane magnetization component of the sample. The assignment of the periodic magnetic contrast modulation to a spin spiral state is confirmed by measurements with an in-plane magnetized tip, see Fig.\,\ref{Fig2}(d): the period of the magnetic contrast on the dislocation lines is the same as in the measurement with out-of-plane tip, Fig.\,\ref{Fig2}(b), but now the amplitude of the contrast differs for different directions. This is a result of the in-plane tip magnetization, which gives different projections of the in-plane spins of the spiral, depending on its propagation direction. Taking into account that there is a non-negligible DM interaction for the Fe/Re interface~\cite{Palotas2014} , we assume in the following that the spin spiral has unique rotational sense and is of cycloidal nature, which is the type of spin rotation that is favored for this symmetry~\cite{Moriya1960,vonBergmannJPCM2014} . On the pseudomorphic Fe monolayer in Fig.\,\ref{Fig2}(d), areas with hexagonal magnetic contrast are observed, corresponding to the two possible N\'{e}el domains (cf.~Fig.\,\ref{Fig2}(a)), as indicated by white circles.

A close analysis shows that the three directions of the dislocation lines not only show a different magnetic contrast amplitude, see also magnified views in Fig.\,\ref{Fig2}(e), but also the adjacent strands within a given dislocation line are different; while two pairs of strands show a magnetic period that is in phase, for the third dislocation line the magnetic contrast on the strands is in antiphase, cf.\ dashed lines. This is reminiscent of recent findings for a spin spiral interacting with similar dislocation lines for Fe layers on Ir$(111)$~\cite{HsuPRL2016} , and originates from a canted spin structure within the spin spiral, see sketched in Fig.\,\ref{Fig2}(f). This model shows a spin structure that is in agreement with both measurements using out-of-plane as well as in-plane sensitive tips, Fig.\,\ref{Fig2}(b)-(e). A tip magnetization direction as indicated in Fig.\,\ref{Fig2}(g) does not only qualitatively reproduce the amplitudes of both strands for each of the three dislocation lines, but in particular also yields identical phases between adjacent strands (dashed lines), in agreement with the experimental data and SP-STM simulations of Fig.\,\ref{Fig2}(e) and (g), respectively. Such a spin canting has been explained previously by a coupling of the spin spiral propagation direction to the bcc [001] directions~\cite{HsuPRL2016} , see the lines of atoms in the structure model of Fig.\,\ref{Fig1}(d), which form roughly an angle of $\pm 25^{\circ}$ with respect to the $<11\overline{2}>$ direction in our system.

\begin{figure}[]
	\includegraphics[width=8.5cm]{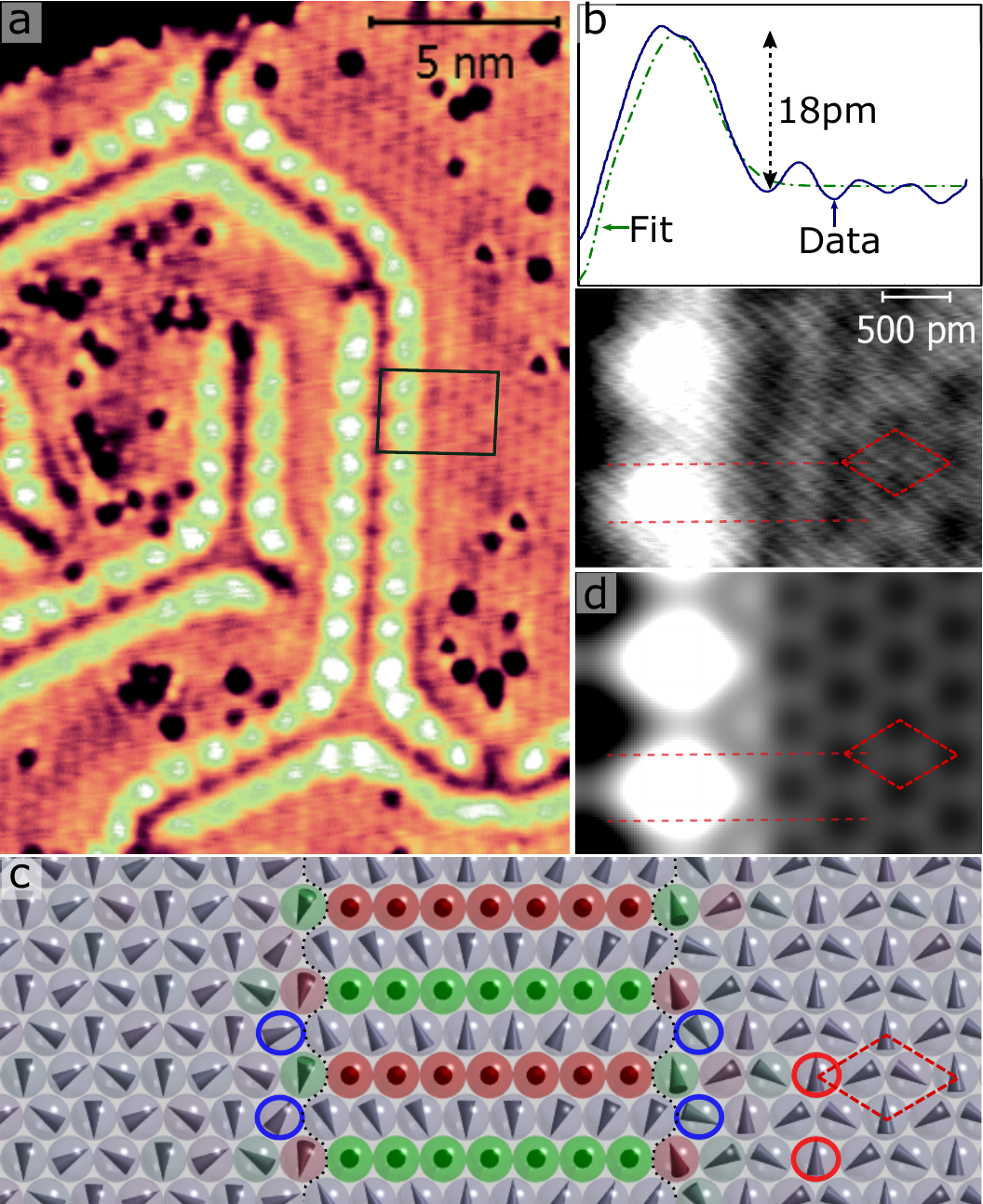}
	\caption{\label{Fig3} (Color online) \textbf{(a)}~SP-STM constant-current image obtained with in-plane sensitive Cr tip on an area where spin spiral and N\'eel states coexist (measurement parameters: $U= +10$\,mV, $I= 1$\,nA, $T=4.6$\,K). \textbf{(b)}~High-resolution constant-current image at the interface between the spin spiral and the N\'eel states, the area is indicated by the rectangle in (a). The unit cell of the N\'eel state is indicated by the diamond. The dashed lines show the correspondence between out-of-plane components of the spin spiral and the N\'eel state. The height profile at the top results from an average over the entire image, the fit consists of three Gaussians and is used for the SP-STM simulations of Fig.\,\ref{Fig2}(g) and Fig.\,\ref{Fig3}(d). \textbf{(c)}~Representative example of Monte-Carlo simulations performed for a single dislocation line embedded in a pseudomorphic area. The color indicates the out-of-plane magnetization components (red up; green down). \textbf{(d)}~SP-STM simulation of the Monte-Carlo result shown in the right part of (c), performed as the one for Fig.\,\ref{Fig2}(g).}
\end{figure}

The question arises, whether there is a coupling between the two coexisting magnetic states, i.e.\ the N\'{e}el state in the pseudomorphic area and the spin spiral following the dislocation lines. The SP-STM image of Fig.\,\ref{Fig3}(a) shows a magnified view of the lower right area of Fig.\,\ref{Fig2}(d), imaged with the same in-plane sensitive tip. The black rectangle indicates an area where the N\'{e}el state and the spin spiral seem to make a sharp transition, and a high-resolution image of this area is displayed in Fig.\,\ref{Fig3}(b). Darker dots, located at the corners of the inserted diamond, indicate atoms with a spin direction antiparallel to the tip. The N\'{e}el state is clearly visible up to about 1.8\,nm from the center of the dislocation line, and it becomes evident that the spin spiral period is indeed exactly twice of the period of the N\'{e}el state in this direction, i.e.\ exactly 4 atomic rows along $[11\overline{2}]$. We find that the atoms with a spin parallel to the dislocation line and 1.8\,nm away from its center (see left corner of the diamond) are always in-phase with the out-of-plane magnetization component of the spin spiral, see dashed lines in Fig.\,\ref{Fig3}(b).

To further investigate the coupling we have performed Monte-Carlo simulations in which the spins in the pseudomorphic area are relaxed from a random state at 80\,K and cooled down to 1\,K. We used an antiferromagnetic nearest neighbor coupling of $J=-15$\,meV \cite{Hardrat2009} . To obtain the same N\'{e}el states as predicted by density functional theory~\cite{Palotas2014} we have introduced an in-plane anisotropy term $\sin ^6 (\theta) \cos (6 \varphi)$~ \cite{Chikazumi1997,vonBergmannNanoLett2015b} with 1\,meV/atom, which favors an alignment of the spins along $<11\overline{2}>$. The DMI was set to $-2$\,meV/atom being $~15\%$ of the isotropic exchange interaction $J$ \cite{Palotas2014} . For the spins being part of the spin spiral guided by the dislocation line the magnetic interaction parameters are expected to be different, because the number of nearest neighbors and their distance, the adsorption site, and the hybridization with the substrate changes. Since these parameters are unknown we fix the spins within the dislocation line to the four atom period with canted spins, see Fig.\,\ref{Fig2}(f). One representative state is shown in Fig.\,\ref{Fig3}(c), where the dotted line indicates the border between fixed spins in the dislocation line and free spins for the pseudomorphic layer. We find both possible N\'{e}el states (see Fig.\,\ref{Fig2}(a)) with equal probability, and a rather sharp transition between the two non-collinear magnetic states, in agreement with the experiment. There are three possible positions of the N\'{e}el state with respect to the dislocation line center: the spins parallel to the dislocation line can either be located at the positions as in the displayed image, i.e.\ 4 atoms away from the last out-of-plane fixed spin, see red circles; or the N\'{e}el state could be laterally phase shifted by one or two atomic distances with respect to the displayed state. The Monte Carlo simulations show that the displayed state has the lowest energy among the three, thereby showing that there is indeed a coupling between the two coexisting non-collinear states. An SP-STM simulation of the right part of Fig.\,\ref{Fig3}(c) is displayed in Fig.\,\ref{Fig3}(d). We observe good agreement with the experimental data, Fig.\,\ref{Fig3}(b), regarding the width of the transition region and the lateral correlation between the spin spiral and the N\'{e}el state.

In the Monte Carlo simulations one can see that the N\'{e}el state is nearly undistorted up to one atomic row next to the fixed spins at the dislocation line. We find that the spins indicated by blue circles are also close to the respective N\'{e}el state, with an angle of $\pm (15-20)^{\circ}$ off from the perfect N\'{e}el state direction. This is the case for both N\'{e}el domains, and we find that both are degenerate. It is worth mentioning that it is rather unlikely that the magnetic interactions in the region of the dislocation line have a spin spiral ground state with a period of exactly four atoms. Instead this period may be imposed by the N\'{e}el state, thus gaining energy due to a favorable coupling between the two non-collinear magnetic states at the cost of a slightly modified spin spiral period in the dislocation line.

In summary, our SP-STM measurements reveal two coexisting non-collinear spin states in the Fe ML on Re(0001): while the magnetic ground state of the pseudomorphic Fe layer is the N\'eel state, the spin state within dislocation lines is a spin spiral with a period about $1.0$~nm. We find that within the spin spiral the spins are canted by about $\pm 25^{\circ}$ with respect to the dislocation line. Monte Carlo simulations reproduce the experimentally observed sharp transition and the coupling between the two different magnetic states: the position of the N\'eel state with respect to the spin spiral is always the same one, and the other two possible phase-shifted domains have a higher energy. It remains to be explored whether the commensurability of the spin spiral is enforced by its coupling to the N\'eel state. 

\section{AUTHOR INFORMATION}
\textbf{Corresponding Authors}

$^{*}$E-mail: kubetzka@physnet.uni-hamburg.de

$^{\dagger}$E-mail: kbergman@physnet.uni-hamburg.de
	
\textbf{Author Contributions}

A.P.M. performed the experiments; A.K. performed the MC calculations and SP-STM simulations; A.P.M. and K.v.B analyzed the data and wrote the manuscript; all authors discussed the manuscript.

\textbf{Notes}

The authors declare no competing financial interest.
\section{ACKNOWLEDGEMENTS}
\begin{acknowledgements}

We thank E.Y.~Vedmedenko and K.~Palotas for discussions. Financial support from the German Research Foundation (DFG) via SFB668-A8 and from the ERC Advanced Grant ASTONISH is gratefully acknowledged. A.~Palacio-Morales acknowledges support by the Alexander von Humboldt foundation.

\end{acknowledgements}


\bibliography{Fe_Re0001_biblio_NanoLett}

\end{document}